\documentclass[12pt]{article}
\pdfoutput=1
\usepackage{putex}
\usepackage[vcentermath]{youngtab}
\usepackage{subfig}
\usepackage{lscape}

\usepackage{graphicx}
\usepackage{epstopdf}
\usepackage{enumerate}
\usepackage{cite}
\usepackage{tensor}
\usepackage{slashed}
\usepackage{amsmath}
\usepackage{amssymb}
\usepackage{mathrsfs}
\usepackage{lgrind}
\usepackage{adjustbox}
\usepackage{hyperref}
\usepackage{tikz}
\tikzset{
    Witten diagram/.style={
        execute at begin picture={
            \draw[blue, line width=1.5pt] circle[radius=\pgfkeysvalueof{/tikz/Witten/radius}];
            \path node (X){\phantom{X}};
        },
        baseline={(X.base)}
    },
    vertex/.style={circle,fill,inner sep=1.5pt,node contents={}},
    Witten/.cd,
    radius/.initial=3cm
}
\usetikzlibrary{patterns}

\numberwithin{equation}{section}

\newcommand {\be} {\begin {equation}}
\newcommand {\ee} {\end {equation}}

\newcommand {\bes} {\begin {equation*}}
\newcommand {\ees} {\end {equation*}}

\def\CD{{\cal D}}

\def\CH{{\cal H}}

\def\CO{{\cal O}}

\newcommand{\beq}{\begin{equation}}
\newcommand{\eeq}{\end{equation}}

\def\be{ \begin{equation} }
\def\ee{ \end{equation} }

\institution{PU}{Joseph Henry Laboratories, Princeton University, Princeton, NJ 08544, USA}
\institution{PCTS}{Princeton Center for Theoretical Science, Princeton University, Princeton, NJ 08544, USA}
\institution{PGI}{Princeton Gravity Initiative, Princeton University, Princeton, NJ 08544, USA}

\begin{document}

\title{
Ginzburg-Landau description of a class of non-unitary minimal models 
}

\authors{Andrei Katsevich,\worksat{\PU} Igor R.~Klebanov\worksat{\PU, \PCTS} and Zimo Sun\worksat{\PU, \PGI}
}

\abstract{
It has been proposed that the Ginzburg-Landau description of the non-unitary conformal minimal model $M(3,8)$ is provided by the Euclidean theory of two real scalar fields with third-order interactions that have imaginary coefficients. The same lagrangian describes the non-unitary model $M(3,10)$, which is a product of two Yang-Lee theories $M(2,5)$, and the Renormalization Group flow from it to $M(3,8)$. This proposal has recently passed an important consistency check, due to Y. Nakayama and T. Tanaka, based on the anomaly matching for non-invertible topological lines. In this paper, we elaborate the earlier proposal and argue that the two-field theory describes the $D$ series modular invariants of both $M(3,8)$ and $M(3,10)$. We further propose the Ginzburg-Landau descriptions of the entire class of $D$ series minimal models $M(q, 3q-1)$ and $M(q, 3q+1)$, with odd integer $q$. They involve $\mathcal{PT}$-symmetric theories of two scalar fields with interactions of order $q$ multiplied by imaginary coupling constants.  
 }

\date{}
\maketitle

\section{Introduction}

The minimal models $M(p,q)$, where $p$ and $q$ are relatively prime positive integers \cite{Belavin:1984vu}, are a famous class of 2D conformal field theories. For over forty years, they have been playing an important role in the physics of various critical phenomena and in string theory. Their central charges are
$c(p,q)= 1- 6 \frac{(p-q)^2}{pq}$, and they possess $\frac{(p-1)(q-1)}{2}$ Virasoro primary operators $\phi_{r,s}$ of holomorphic dimensions 
\be
h_{r,s}=\frac{(rq-sp)^2-(p-q)^2}{4pq}\ ,  \quad r=1, \ldots, p-1;\ s=1, \ldots q-1\ .
\ee
The primary operators are identified according to $\phi_{r,s}\equiv \phi_{p-r, q-s}$. 

Among the minimal models, a distinguished subset are the unitary models $M(m+1, m+2)$ \cite{Friedan:1983xq}, where $m=2, 3, \ldots$. For $m=2$ this is the Ising model, for $m=3$ the tricritical Ising, for $m=4$ the 3-state Potts, and so on. 
Zamolodchikov \cite{Zamolodchikov:1986db} has proposed that the Ginzburg-Landau (GL) effective description of $M(m+1, m+2)$ is given by the Euclidean scalar field theory
\be
\label{ZamGL}
S_{m+1, m+2}=\int d^d x \bigg ( \frac{1}{2}\left(\partial_{\mu}\phi\right)^2 + {g\over (2m)!} \phi^{2m} 
\bigg )\ ,
\ee
where we assume that all the coefficients of terms $\phi^{2k}$ with $k<m$ are tuned to zero.
In $d=2$, the coupling $g$ has dimension of mass-squared, so these models flow to strong coupling. However, they become weakly coupled near the upper critical dimensions $d_c(m)=\frac{2m}{m-1}$, and one can develop the $d_c-\epsilon$ expansions. For example, for the $\phi^4$ theory ($m=2$) Wilson and Fisher \cite{Wilson:1971dc} developed the $4-\epsilon$ expansions that, when continued to $d=2$, give good approximations to the $M(3,4)$ scaling dimensions: $\Delta_\phi=2h_{2,2}=1/8$ and 
$\Delta_{\phi^2}=2h_{2,1}=1$. In general, $\phi$ is identified with the primary operator $\phi_{2,2}$ which has the lowest positive holomorphic dimension, $h_{2,2}= \frac{3}{4(m+1)(m+2)}$.

Currently, there is growing interest in the non-unitary minimal models $M(p,q)$, where $|p-q|>1$. They may have various applications, including the nearly scale invariant regime of turbulence in $2+1$ dimensions 
\cite{Polyakov:1992er,falkovich2015operator}.
Only some partial results about the GL description of the non-unitary models are available so far.
An interesting class of non-unitary models are $M(2, 2 m+1)$ with $m=2, 3, \ldots$. The first representative is $M(2,5)$ corresponding to the Yang-Lee edge singularity of the Ising model with an imaginary magnetic field. Its GL description is provided by the scalar field theory with interaction $\sim i \phi^3$ \cite{Fisher:1978pf,Cardy:1985yy}. 
Indeed, the theory
\be
\label{FishGL}
S_{2,5}=\int d^d x \bigg ( \frac{1}{2}\left(\partial_{\mu}\phi\right)^2 + {g\over 6} \phi^{3} 
\bigg )\ ,
\ee    
has a weakly coupled IR fixed point in $d=6-\epsilon$ with an imaginary $g$. The $i\phi^3$ theory is an example of $\mathcal{PT}$-symmetric Euclidean field theory
(see for example \cite{Bender:2018pbv}); the symmetry acts by $\phi\rightarrow -\phi$, $i\rightarrow -i$.\footnote{This is not a conventional 
$\mathcal{PT}$ symmetry, since it includes an internal transformation $\phi\rightarrow -\phi$.}
 The imaginary coupling constant removes the instability in the functional integral, although it oscillates rapidly. The $6-\epsilon$ expansion 
 can be used to approximate the observables \cite{Fisher:1978pf}.  
For example, the expansion of $\Delta_\phi$ is now known to order $\epsilon^5$, and its various  
Pad\' {e} extrapolations were carried out in \cite{Kompaniets:2021hwg,Borinsky:2021jdb} and compared with the exact 2D value $2h_{1,2}=-\frac{2}{5}$.

The next $\mathcal{PT}$-symmetric odd potential is $\sim i\phi^5$, and it was conjectured to describe the $(2,7)$ minimal model 
\cite{Katsevich:2025ojk} (for earlier work, see
\cite{amslaurea11308,Zambelli:2016cbw,Lencses:2022ira,Lencses:2024wib}).
It appears that all the theories
\be
\label{oddGL}
S_{2, 2m+1}=\int d^d x \bigg ( \frac{1}{2}\left(\partial_{\mu}\phi\right)^2 + {g\over (2m-1)!} \phi^{2m-1} 
\bigg )\ ,
\ee   
where $m$ is a positive integer, possess IR fixed points in $2 \frac{2m-1}{2m-3} - \epsilon$ dimensions at imaginary $g$ and are, therefore, $\mathcal{PT}$-symmetric. 
They were conjectured to provide GL descriptions of the $M(2, 2m+1)$ models in 2D \cite{Katsevich:2025ojk}. 
The Virasoro primary operators in the minimal model are $\phi_{1, k}$, $k=1, \ldots m$. The GL field $\phi$ appears to correspond  to $\phi_{1,2}$, the operator of the 
least negative holomorphic dimension: $h_{1,2}=-\frac{m-1}{2m+1}$.

Building on these results, it is of obvious interest to look for the 2D minimal models described by GL theories containing two scalar fields. A well-known 
unitary minimal model with such a description is the $3$-state Potts model, which is the $D$ series modular invariant of $M(5,6)$ \cite{Cappelli:1987xt,Kato:1987td}. 
It corresponds to a GL theory of one complex field $\rho$ with potential of the $\mathbb{Z}_3$ invariant form $g_1 (\rho^3+\bar \rho^3)+ g_2(\rho\bar \rho)^2$ \cite{Zia:1975ha,Amit_1979}. In general, the $D$ series modular invariant \cite{Cappelli:1987xt,Kato:1987td} is obtained by orbifolding the $\mathbb Z_2$ symmetry in the $A$ series modular invariant \footnote{There is a $\mathbb Z_2$ symmetry in the $A$ series modular invariant unless $p$ or $q$ is equal to 2 \cite{Lassig:1991an}. When $p$ and $q$ are both odd, the $\mathbb Z_2$ symmetry is anomalous 
\cite{Nakayama:2022svf,Cordova:2019wpi}
and there is no $D$ series modular invariant.}. The $\mathbb Z_2$ orbifold keeps the even states, but also adds the $\mathbb Z_2$ twisted sector \cite{DiFrancesco:1997nk}. As a result, a new $\mathbb Z_2$ symmetry emerges in the $D$ series modular invariant which identifies the twisted sector as $\mathbb Z_2$ odd.

In \cite{Fei:2014xta,Klebanov:2022syt} it was argued that the non-unitary models $M(3,10)$ and $M(3,8)$ are described by different fixed points of the following theory of two scalar fields with cubic interactions: 
\begin{equation} \label{eq:GL_action}
S_{3} = \int d^dx \left( \frac{1}{2} (\partial_\mu \phi )^2+\frac{1}{2} (\partial_\mu \sigma )^2+\frac{g_1}{2} \sigma \phi ^2+\frac{g_2}{6} \sigma ^3\right)\ .
\end{equation}  
A new powerful argument in favor of this identification was recently provided by Nakayama and Tanaka \cite{Nakayama:2024msv} using the topological defect line anomaly matching 
\cite{Chang:2018iay}. This led to the proposal of
an infinite set of non-unitary RG flows \cite{Nakayama:2024msv}\footnote{The operator $\phi_{1,2k+1}$ can generally mix along the RG flow with the more relevant operators
$\phi_{1,2l+1}$, $l<k$, so that some fine tuning is needed to reach the $M(kq-I, q)$ fixed point \cite{Nakayama:2024msv}.}
\be
\label{kflow}
M(kq+I, q) + \phi_{1,2k+1}\rightarrow  M(kq-I, q) \ .
\ee
The $k=1$ and $k=2$ families of such flows were studied in the earlier literature 
\cite{Zamolodchikov:1987ti,Ahn:1992qi,Lassig:1991an,Martins:1992yk,Ravanini:1994pt,Dorey:2000zb}, while
the flow from $M(3,10)$ to $M(3,8)$ \cite{Fei:2014xta,Klebanov:2022syt} is a representative of the $k=3$ family with $q=3$, $I=1$.
The families with half-integer $k$ are allowed as well \cite{Nakayama:2024msv}; in particular, the $k=\frac{1}{2}$, $I=\frac{1}{2}$ flows were described in \cite{Martins:1992yk,Ravanini:1994pt,Dorey:2000zb}.

In spite of the progress during the past few years \cite{amslaurea11308,Zambelli:2016cbw,Fei:2014xta,Klebanov:2022syt,Lencses:2022ira,Lencses:2024wib}, the GL descriptions of most non-unitary minimal models remain unknown. In this paper we make some progress in this direction. We generalize the GL description of the minimal models $M(3,8)$ and $M(3,10)$ \cite{Fei:2014xta,Klebanov:2022syt} to the whole class of models $M(q, 3q-1)$ and $M(q, 3q+1)$ with odd integer $q$. It is provided by the effective action (\ref{eq:GLgen_action}) containing two scalar fields with imaginary interactions of order $q$. The RG flow connecting the two models belongs to the general class recently described in \cite{Nakayama:2024msv}: it is an example of (\ref{kflow}) with $k=3$ and $I=1$. Since our GL theory contains two scalar fields, the operator spectrum includes the spin-$1$ field
\be
\label{nccurrent}
J_\mu= \sigma \partial_\mu \phi- \phi \partial_\mu \sigma\ .
\ee
Therefore, such a GL theory cannot correspond to an $A$ series modular invariant, which contains only the spin-$0$ primary fields. We will be matching the two-field GL theories with the $D$ series modular invariants, which exist for $M(p,q)$ when either $p$ or $q$ is even and greater than $4$ \cite{Cappelli:1987xt,Kato:1987td}. We will discuss the special cases $M(3,8)$, $M(3,10)$, $M(5,14)$ and $M(5,16)$ in some detail.

\section{GL description of $M(3,10)$ and $M(3,8)$}

In \cite{Fei:2014xta,Klebanov:2022syt} it was argued that the non-unitary models $M(3,10)$ and $M(3,8)$ are described by different fixed points of GL action (\ref{eq:GL_action}) in $6-\epsilon$ dimensions. This action is the special case $N=1$ of the $O(N)$ invariant theories introduced in \cite{Fei:2014yja} in search of UV completions of the quartic $O(N)$ invariant theories for dimensions $4<d<6$. 
For $N=1$, the two coupling constants take imaginary values, and the fixed points have $\mathcal{PT}$ symmetry under $\sigma\rightarrow -\sigma$, $i\rightarrow -i$, in addition to the
$\mathbb{Z}_2$ symmetry under $\phi\rightarrow -\phi$ \footnote{Unlike in the  $S_{2, 4m+1}$ theories, there is an ambiguity in defining $\mathcal{PT}$ due to the $\mathbb{Z}_2$ symmetry. For a given choice of  $\mathcal{PT}$, its product with the $\mathbb Z_2$ transformation defines a new $\mathcal{PT}$ symmetry.}. At one of the fixed points, $g_1^*=g_2^*$, one finds two decoupled Yang-Lee theories (\ref{FishGL}) for fields
$\phi_1=(\sigma+\phi)/\sqrt 2$ and $\phi_2=(\sigma-\phi)/\sqrt 2$,
\be
S_{3,10}=\int d^d x \bigg ( \frac{1}{2}\left(\partial_{\mu}\phi_1\right)^2 + \frac{1}{2}\left(\partial_{\mu}\phi_2\right)^2 +
{g\over 6} \left (\phi_1^{3} +  \phi_2^{3}\right )
\bigg )\ .
\ee    
 In 2D the product of these two models corresponds to $M(3,10)$ with the $D_6$ modular invariant
\cite{Kausch:1996vq,Quella:2006de,2011NJPh...13d5006A}. The $D_6$ partition function is given by 
\begin{align}
Z^{D_6}_{3,10}=|\chi_{1,1}+\chi_{1,9}|^2+ |\chi_{1,3}+\chi_{1,7}|^2 + 2|\chi_{1,5}|^2.
\end{align}

\begin{table}[t]
        \centering
          \renewcommand{\arraystretch}{1.2}
        \begin{tabular}{|c|c|c|c|c|c|c|}
            \hline
            $(3,10)$ & $\phi_{1,1}$ &  $\phi_{1,3}$ & $\phi^{+}_{1,5}$ & $\phi^{-}_{1,5}$ & $\phi_{1,7}$ & $\phi_{1,9}$ \\\hline
            $h$ & $0$ & $-\frac{2}{5}$ & $-\frac{1}{5}$ & $-\frac{1}{5}$ & $\frac{3}{5}$ & $2$ \\\hline
            $\mathbb{Z}_2$ & even & even & even & odd & even & even  \\\hline
            $\mathcal{PT}$ & even & even & odd & even & even & even  \\\hline
            GL & 1 & $\phi_1\phi_2$ & $\phi_1+\phi_2$ & $\phi_1-\phi_2$ & $i \phi_1\phi_2(\phi_1+\phi_2)$ & $T_{1\mu \nu} T_2^{\mu \nu}$ \\\hline
        \end{tabular}
        \caption{Spin-$0$ primary operators and their properties for the $D_6$ modular invariant of $M(3,10)$.}
        \label{tab:M310}
    \end{table}

The matching of this model with the GL description was discussed in \cite{Klebanov:2022syt}, and we review it here. We denote the holomorphic dimension by $h$ and the antiholomorphic dimension by 
$\bar h$.
There are two scalar primary operators with $(h, \bar h)=(-\frac{1}{5}, -\frac{1}{5})$, 
\be
\label{onefive}
\phi_{1,5}^+= \phi_1+ \phi_2\sim \sigma\ , \qquad  \phi_{1,5}^-= \phi_1- \phi_2 \sim \phi\ .
\ee
The first of them is even under the $\mathbb{Z}_2$ symmetry, and the second is odd.
The operator $\phi_{1,3}$ with $(h, \bar h)=(-\frac{2}{5}, -\frac{2}{5})$ corresponds to $\phi_1 \phi_2$, and the operator $\phi_{1,7}$ with $(\frac{3}{5}, \frac{3}{5})$ to $i\phi_1 \phi_2 (\phi_1+\phi_2)$. 
The operators $\phi_{1,3} \bar \phi_{1,7}\oplus \bar \phi_{1,7} \phi_{1,3}$ with dimensions $(-\frac{2}{5}, \frac{3}{5})$ and
$(\frac{3}{5}, -\frac{2}{5})$ form a vector corresponding to $\phi_1\partial_\mu \phi_2-  \phi_2\partial_\mu \phi_1$. 
The operators with dimensions
$(2, 0)$ and $(0, 2)$ form a conserved spin-$2$ tensor, corresponding to the {\it difference} of the stress-energy tensors of the two decoupled GL theories, $T_1^{\mu \nu}- T_2^{\mu \nu}$.
Finally the operator $\phi_{1,9}$ with  $(h, \bar h)=(2, 2)$ corresponds to $T_{1\mu \nu} T_2^{\mu \nu}$. In spacetime dimension $d$, its exact scaling dimension is $\Delta=2d$.

Let us consider RG flows originating from this $M(3,10)$ CFT. Its perturbation by the operator $\phi_{1,5}^+$ creates RG flow corresponding to $k=2$, $q=3$, $I=4$ in (\ref{kflow}); it is expected to lead to $M(2,3)$, which is the massive theory. This agrees with the field identification (\ref{onefive}), which implies that the theory is a product of two Yang-Lee models, each one perturbed by its non-trivial primary field $\phi_{1,2}\equiv \phi_{1,3}$. Each one undergoes the $k=2$ flow from $M(2,5)$ to the massive phase \cite{Kausch:1996vq}. 
The perturbation of $M(3,10)$ by $\phi_{1,3}$ also creates RG flow to the massive phase \cite{Katsevich:2024sov}. On the other hand, perturbing $M(3,10)$ by $\phi_{1,7}$ creates $k=3$, $q=3$, $I=1$ flow to $M(3,8)$ 
\cite{Fei:2014xta,Klebanov:2022syt,Nakayama:2024msv}.\footnote{Since operators $\phi_{1,3}$ and $i\phi_{1,5}^+$ have the same quantum numbers as $\phi_{1,7}$, they are expected to be induced along the RG flow as well, so that some fine tuning is needed for the flow to reach $M(3,8)$ \cite{Nakayama:2024msv}. In $d=6-\epsilon$, where $\phi_{1,7}$ becomes nearly marginal, the flow can be studied by integrating the $\beta$ functions calculated using the rules of renormalized perturbation theory. Thus, slightly below the upper critical dimension $6$, the relevant operators $\phi_{1,3}$ and $i\phi_{1,5}^+$ don't need to be included explicitly.}
In $d=2$, this flow obeys the RG inequality \cite{Castro-Alvaredo:2017udm} for the quantity $c_{\rm eff}= c- 24 h_{\rm min}$ introduced in \cite{Itzykson:1986pk} (see also \cite{Seiberg:1990eb}). For the minimal models, $c_{\rm eff} (p, q)=1- \frac{6}{pq}$. 
The inequality $c_{\rm eff}^{\rm UV}> c_{\rm eff}^{\rm IR}$ applies to all the
$\mathcal{PT}$-symmetric RG flows \cite{Castro-Alvaredo:2017udm}, 
and this was one of the arguments for the GL description of $M(3,8)$ \cite{Klebanov:2022syt}.

The $D_5$ modular invariant partition function of $M(3, 8)$ is
\begin{align}
Z^{D_5}_{3,8}=|\chi_{1,1}|^2+|\chi_{1,3}|^2+ |\chi_{1,5}|^2+|\chi_{1,7}|^2 + |\chi_{1,4}|^2 + \chi_{1,2}\bar\chi_{1,6}+\chi_{1,6}\bar\chi_{1,2}~,
\end{align}
where the first $4$ terms are from the untwisted sector (they are the $\mathbb{Z}_2$ even states in the $A$ series modular invariant), while the remaining $3$ from the twisted sector.
There are $5$ scalar primary operators, whose quantum numbers under the discrete symmetries and GL identifications are listed in Table \ref{tab:M38}. 
The $D_5$ modular invariant of $M(3,8)$ also contains the spin-$1$ operator $\phi_{1,2} \bar \phi_{1,6}\oplus \bar \phi_{1,2} \phi_{1,6}$ with dimensions
$(-\frac{7}{32}, \frac{25}{32})$ and $(\frac{25}{32}, -\frac{7}{32})$, 
and we identify it with the GL operator (\ref{nccurrent}).

\begin{table}[t]
        \centering
        \renewcommand{\arraystretch}{1.2}
        \begin{tabular}{|c|c|c|c|c|c|c|c|c|c|}
            \hline
            $(3,8)$ & $\phi_{1,1}$ & $\phi_{1,3}$ & $\phi^-_{1,4}$ & $\phi_{1,5}$ &  $\phi_{1,7}$ \\\hline
            $h$ & $0$ &  $-\frac{1}{4}$ & $-\frac{3}{32}$ & $\frac{1}{4}$ &  $\frac{3}{2}$  \\\hline
            $\mathbb{Z}_2$ & even & even & odd & even & even \\\hline
            $\mathcal{PT}$ & even & odd & even & odd &  even \\\hline
            GL & 1 &  $\sigma$ & $\phi$ & $i\sigma^2+i\phi^2$  & $i\phi^2\sigma+i\sigma^3$ \\\hline
        \end{tabular}
        \caption{Spin-$0$ primary operators and their properties for the $D_5$ modular invariant of $M(3,8)$. We schematically use the notation $\CO_1+\CO_2$ to denote a linear combination of the two operators $\CO_1$ and $\CO_2$ that is a primary. 
}
        \label{tab:M38}
    \end{table}

\begin{figure}[t]
\centering
\begin{tikzpicture} 
\draw [line width = 0.5mm] (-1, -1) to (0,0.5) node[vertex]{};
\draw [line width = 0.5mm, dashed] (0,0.5)  to (1,-1);
\node at (0,-1.2) {$\mathcal{D}_0$};
\end{tikzpicture}
\qquad 
\begin{tikzpicture} 
\draw [line width = 0.5mm] (-1, -1) to (-0.5,-0.25) to (0, 0.5)node[vertex]{};
\draw [line width = 0.5mm, dashed] (0,0.5)  to (0.5,-0.25) to (1,-1);
\draw [line width = 0.5mm, dashed] (-0.5,-0.25)  to (0.5,-0.25);
\node at (0,-1.2) {$\mathcal{D}_1$};
\end{tikzpicture}
\qquad 
\begin{tikzpicture} 
\draw [line width = 0.5mm] (-1, -1) to (-0.5,-0.25) to (0.5, -0.25) to (0,0.5)node[vertex]{};
\draw [line width = 0.5mm, dashed]  (-0.5,-0.25) to (0,0.5);
\draw [line width = 0.5mm, dashed] (0.5,-0.25)  to (1,-1);
\node at (0,-1.2) {$\mathcal{D}_2$};
\end{tikzpicture}
\qquad 
\begin{tikzpicture} 
\draw [line width = 0.5mm] (-1, -1) to (0,-0.5) to (0, -0.25) to [in=-150, out=150] (0,0.5)node[vertex]{};
\draw [line width = 0.5mm, dashed] (1, -1) to (0,-0.5);
\draw [line width = 0.5mm, dashed]  (0, -0.25) to [in=-30, out=30] (0,0.5);
\node at (0,-1.2) {$\mathcal{D}_3$};
\end{tikzpicture}
\caption{One-loop renormalization of $J_\mu = \sigma\partial_\mu\phi-\phi\partial_\mu\sigma$. The diagrams represent the three-point function $\langle J_\mu \phi(p)\sigma(q)\rangle$. The solid line denotes the $\phi$ propagator and the dashed line denotes the $\sigma$ propagator.}
\label{Jrenormalization}
\end{figure}
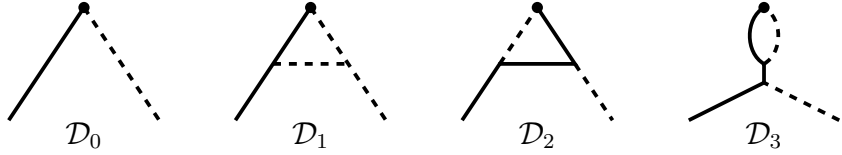

Let us compute the one-loop scaling dimension of $J_\mu$ in $6-\epsilon$ dimensions. There are three diagrams contributing to the renormalization of $J_\mu$ at the one-loop level. They are denoted by $\mathcal{D}_1, \mathcal{D}_2$ and $\mathcal{D}_3$ in figure \ref{Jrenormalization}. The diagram $\CD_3$ vanishes due to a $\mathbb Z_2$ symmetry of the integrand. Both $\CD_1$ and $\CD_2$ are proportional to the integral 
\begin{align}
\int\frac{d^d k}{(2\pi)^d}\frac{p^\mu-q^\mu-2 k^\mu}{k^2(p-k)^2(q+k)^2} =\frac{p^\mu\!-\!q^\mu}{3(4\pi)^3\epsilon}+O(1)~.
\end{align}
Altogether, we find the anomalous dimension of $J_\mu$ to be 
\begin{align}
\gamma_J = \frac{g_1(g_1-g_2)}{3(4\pi)^3}+\gamma_\phi+\gamma_\sigma~,
\end{align}
where $\gamma_\phi = \frac{g_1^2}{6(4\pi)^3}$ and $\gamma_\sigma = \frac{g_1^2+g_2^2}{12(4\pi)^3}$ \cite{Fei:2014yja}. At the IR stable fixed point corresponding to $M(3, 8)$, 
\be
g_2^{*2}=\frac{36}{25} g_1^{*2} = - \frac{216  (4\pi)^3\epsilon}{499}\ ,
\ee
and the scaling dimension of $J_\mu$ becomes 
\begin{align}
\Delta_J = \Delta_\phi+\Delta_\sigma+1 + \frac{10\epsilon}{499} +O(\epsilon^2) = 5-\frac{1089}{998}\epsilon+O(\epsilon^2)~.
\end{align}
Taking $\epsilon=4$ yields $\Delta_J\approx 0.635$ in 2D. The exact value of $\Delta_J$ in 2D is $h_{1,2}+h_{1,6}=\frac{9}{16}\approx 0.563$. Thus, the leading $6-\epsilon$ expansion result is only about $10\%$ off the exact 2D value. 

We note that there is only one $\mathbb Z_2$ odd scalar operator in Table \ref{tab:M38},  $\phi^-_{1,4}$, and it must be identified
with the GL field $\phi$. 
This identification appears to be consistent with the Operator Product Expansions (OPE), which are known for the minimal models \cite{Dotsenko:1984ad}.
In particular, all the terms in the OPE
$\phi^-_{1,4} \times \phi^- _{1,4} \sim 1+i \phi _{1,3} +i \phi _{1,5} +\phi _{1,7}$ can be identified in the GL theory ($i\phi_{1,3}\sim i\sigma$ appears due to a cubic interaction vertex, while $\phi_{1,7}$ due to loop diagrams containing the interaction vertices). 
 Other OPE also appear to be consistent with the operator identifications in Table \ref{tab:M38}. As we will describe later, for the general class $M(q, 3q-1)$ the suggested identification of $\phi$ is with the operator $\phi^-_{(q-1)/2,(3q-1)/2}$ coming from the twisted sector of the $\mathbb Z_2$ orbifold. 

For some scalar operators in Table \ref{tab:M38}, the $6-\epsilon$ expansions provide additional support for the operator identifications.
For example, $\Delta_\phi= 2- \frac{549}{998} \epsilon + O(\epsilon^2)$ \cite{Fei:2014xta,Klebanov:2022syt}, and substituting $\epsilon=4$ gives the 2D estimate $\Delta_\phi\approx -0.2$. This is very close to the exact value $2 h_{1,4} = -\frac{3}{16}\approx -0.19$.\footnote{  
The dimension of $\phi$ is known to order $\epsilon^5$, and its Pad\'{e} extrapolation \cite{Klebanov:2022syt} yielded values further from the exact result; however, they exhibited considerable dependence on how many orders were included.} The Pad\'{e} extrapolation of $\Delta_\sigma$ agrees quite well \cite{Klebanov:2022syt} with the exact 2D dimension $2 h_{1,3}=-\frac{1}{2}$.

Let us note that one combination of $\phi^2$ and $\sigma^2$ is a descendant, while the other corresponds to $\phi_{1,5}$. Perturbing $M(3,8)$ by this operator causes the $k=2$ flow to $M(3,4)$ \cite{Martins:1992yk,Ravanini:1994pt,Dorey:2000zb,Nakayama:2024msv}. This can be understood heuristically via integrating out the massive field $\sigma$ to generate the $\phi^4$ potential in the infrared. A more precise description of the flow $M(3,8)+ \phi_{1,5}\rightarrow M(3,4)$ shows that $c_{\rm eff}$ is not monotonic \cite{Dorey:2000zb}.

\section{Generalization to $M(q, 3q \pm 1)$ with odd $q$}

In \cite{Fei:2014xta} it was noted that $M(2,5)$ and $M(3,8)$ belong to the family of non-unitary minimal models $M(q, 3q-1)$. In this section, we propose that 
the $D_{(3q+1)/2}$ modular invariants of models $M(q, 3q-1)$
 with odd integer $q$ have GL descriptions using two scalar fields with interactions of order $q$:
\begin{equation} \label{eq:GLgen_action}
S_{q} = \int d^dx \left( \frac{1}{2} (\partial_\mu \phi )^2+\frac{1}{2} (\partial_\mu \sigma )^2+\frac{g_1}{(q-1)!} \sigma \phi^{q-1}+\frac{g_2}{6 (q-3)!} \sigma^3\phi^{q-3}
+\ldots + \frac{g_{(q+1)/2}}{q!} \sigma^{q}
\right),
\end{equation}
where all $g_i$ are imaginary. We conjecture that a different imaginary fixed point of the same theory describes the $D_{(3q+3)/2}$ modular invariant of $M(q,3q+1)$. The upper critical dimension of this scalar theory is 
$d_c(q)=2\frac{q}{q-2}$, and we have included all the marginal interaction terms consistent with the discrete symmetries, so that 
the theory is renormalizable in the upper critical dimension.
The fixed points in $d_c(q)-\epsilon$ dimensions can be studied using perturbation theory \cite{Gracey:2017okb}, and so can the RG flows of the coupling constants $g_i$ connecting the different fixed points.

Let us consider the family of $\mathcal{PT}$-symmetric RG flows
\be
\label{threeflow}
M(3q+1, q) + \phi_{1,7} \rightarrow  M(3q-1, q)  
\ee
with odd $q$, which correspond to $k=3$, $I=1$ in (\ref{kflow}). These are generalizations of the $q=3$ flow studied in  \cite{Fei:2014xta,Klebanov:2022syt}, and again the operator 
$\phi_{1,7}$ corresponds to a combination of the interaction terms present in the GL action.
Upon dimensional continuation from the upper critical dimension to $d=2$, the difference of central charges becomes small for large $q$ (in this limit, $c(q,3q \pm 1)\rightarrow -7$):
\be
c_{\rm UV}-  c_{\rm IR}= 
-\frac{12 (8 q^2-1)}{q (9 q^2-1)}=- \frac{32}{3 q} + O(q^{-3})\ , 
\ee
but not as small as in the unitary flows $M(m+1, m+2)+ \phi_{1,3} \rightarrow M(m, m+1)$ where $\delta c\sim 1/m^3$. This is related to the fact that the 3-pt function of 
the perturbing operator $\phi_{1,7}$ falls off as $1/q$ at large $q$ \cite{Dotsenko:1984ad}. As a result, it appears that the flows (\ref{threeflow}) cannot be studied using the 
conformal perturbation theory at large $q$ in a parametrically reliable way. 
The negativity of $c_{\rm UV}-  c_{\rm IR}$ is a sign of the non-unitarity of the theories.
In the UV theory $M(q, 3q+1)$, the RG flow is sourced by the $\mathbb{Z}_2$ even operator $\phi_{1,7}$ of scaling dimension 
\be
\Delta_{\rm UV}=2 h_{1,7}=\frac{6(q-1)}{3q+1}= 2 - \frac{8}{3q} + O(q^{-2})\ , 
\ee
which for large $q$ becomes slightly relevant. In the IR theory $M(q, 3q-1)$, the flow ends with its $\mathbb{Z}_2$ even operator $\phi_{1,7}$ of scaling dimension 
\be
\Delta_{\rm IR}=\frac{6(q+1)}{3q-1}= 2 + \frac{8}{3q} + O(q^{-2})\ . 
\ee

Some of these observations apply to the more general flows (\ref{kflow}). The difference of central charges is
\be
c_{\rm UV}-  c_{\rm IR}= 
-\frac{12 I (q^2(k^2-1)- I^2)}{q ( q^2 k^2- I^2)}=- \frac{12 I (k^2-1) }{k^2 q} + O(q^{-3})\ .
\ee
Since for $k>1$ this scales as $1/q$ for large $q$, the conformal perturbation theory is not expected to be applicable even in this limit. For $k=2$ flows, this observation was made already in
\cite{Ravanini:1994pt}.
In the UV theory $M(q, kq+I)$, the RG flow is sourced by the $\mathbb{Z}_2$ even operator $\phi_{1,2k+1}$ of scaling dimension 
\be
\Delta_{\rm UV}=2 h_{1,2k+1}=\frac{2 k(q-I)}{k q+I}= 2 - 2\frac{(1+k) I}{k q} + O(q^{-2})\ , 
\ee
which for large $q$ becomes slightly relevant. In the IR theory $M(q, kq-I)$, the RG flow ends with the $\mathbb{Z}_2$ even operator $\phi_{1,2k+1}$ of scaling dimension 
\be
\Delta_{\rm IR}=\frac{2 k(q+I)}{k q-I}= 2 +2\frac{(1+k) I}{k q} + O(q^{-2})\ , 
\ee

Now let us discuss operator identifications in the GL description (\ref{eq:GLgen_action}) of $M(q, 3q\pm 1)$.
We identify the $\mathbb{Z}_2$ even operator $\phi_{1,7}$ 
with GL operator of the form $i\sum_{j=1}^{\frac{q+1}{2}}\sigma^{2j-1}\phi^{q-2j+1}$, i.e. the terms in the GL potential with different coefficients. 
In $M(q, 3q - 1)$, 
we identify the GL field $\sigma$ with $\phi_{1,3}$, which is the scalar operator with minimum dimension $h_{1,3}= - \frac{q-1}{3q- 1}$. 
We identify the operator $\phi_{1,5}$, which is $\mathcal{PT}$ odd, with a GL operator that is a mixture of $i\sigma^2$ and $i\phi^2$.
Its dimension is $h_{1,5}= \frac{1}{3q-1}$.
According to \cite{Martins:1992yk,Ravanini:1994pt,Dorey:2000zb,Nakayama:2024msv}, 
there is $k=2, I=q-1$ RG flow $M(3q-1, q) \rightarrow  M(q+1, q)$. It should be sourced by the $\mathcal{PT}$ even operator  $i\phi_{1,5}$.
In the GL description, the imaginary coefficient of $\phi_{1,5}$ is important because it makes the $\sigma$ field massive. 
Then, integrating out $\sigma$ produces the effective potential $\phi^{2q-2}$ for the remaining field $\phi$, while the lower powers of $\phi$ need to be tuned away.

In $M(q, 3q + 1)$, we propose to identify the $\mathbb Z_2$ even primary operator $\phi_{1,5}$ of dimension $h_{1,5}= -\frac{2}{3q+1}$ with the GL field $\sigma$. We then identify  $\phi_{1,3}$ of dimension $h_{1,3}= - \frac{q+1}{3q+1}$ with a mixture of $\sigma^2$ and
$\phi^2$. These identifications are motivated by the OPE
\begin{align}
\phi_{1,5}\times \phi_{1,5} \sim 1+\phi_{1,3}+i\phi_{1, 5}+\phi_{1, 7}+\phi_{1, 9}~.
\end{align}

The $D$ series modular invariant of $M(q, 3q\pm 1)$ contains $\frac{q-1}{2}$ $\mathbb Z_2$ odd scalar primary operators: $\phi^-_{n, \frac{3q\pm 1}{2}}, n=1, 2, \cdots, \frac{q-1}{2}$, among which only the $n= \frac{q-1}{2}$ operator has a negative scaling dimension 
\begin{align}
h_{\frac{q-1}{2}, \frac{3q\pm 1}{2}} = -  \frac{(7q\pm 3)(q\pm 1)}{16q(3q\pm 1)} = - \frac{7}{48} + O(q^{-1})~.
\end{align}
We propose to identify the GL field $\phi$ with $\phi^-_{\frac{q-1}{2}, \frac{3q\pm 1}{2}}$. It is the scalar operator whose negative dimension is the closest to zero. 
The $D$ series modular invariant contains the unique spin-$1$ operator 
\be
\phi_{\frac{q-1}{2},\frac{3q\pm 1}{2}-2} \bar\phi_{\frac{q-1}{2},\frac{3q\pm 1}{2}+2}\oplus \phi_{\frac{q-1}{2},\frac{3q\pm 1}{2}+2} \bar\phi_{\frac{q-1}{2},\frac{3q\pm 1}{2}-2}
\ .
\ee
It is $\mathbb Z_2$ odd, and we identify it with the GL current (\ref{nccurrent}).

\subsection{Some examples}

Let us explore the proposed description of $M(5,14)$ and $M(5,16)$ as a GL theory (\ref{eq:GLgen_action}) with two scalar fields and quintic interactions. 
First, we recall that the $\frac{10}{3}-\epsilon$ expansions for other theories with quintic interactions were carried out in 
\cite{Codello:2017epp,Klebanov:2021sos,Katsevich:2025ojk} following \cite{Gracey:2017okb}.
The theory of a single scalar field is
\be
\label{OSpfour}
S_{2,7}=\int d^d x \bigg (
\frac{1}{2}\left(\partial_{\mu}\phi\right)^2 +  \frac{g}{120}  \phi^5   
\bigg ) \ .
\ \ee
In \cite{Katsevich:2025ojk} it was conjectured to correspond to $M(2,7)$. 

Our proposed  GL description of the $D_8$ modular invariant of $M(5,14)$ is
    \begin{align}
\label{fiveaction}
        S_{5}=\int d^dx\left(\frac{1}{2}(\partial_\mu\phi)^2+\frac{1}{2}(\partial_\mu\sigma)^2+\frac{g_1}{4!}\sigma\phi^4+\frac{g_2}{2!\cdot3!}\sigma^3\phi^2+\frac{g_3}{5!}\sigma^5\right),\quad g_i\in i\mathbb{R}.
    \end{align}
The partition function,
\begin{align}
    Z^{D_8}_{5,14}=\sum_{i=1}^2(|\chi_{i,1}+\chi_{i,13}|^2+|\chi_{i,3}+\chi_{i,11}|^2+|\chi_{i,5}+\chi_{i,9}|^2+2|\chi_{i,7}|^2)~,
\end{align}
includes the twisted sector which is odd under the $\mathbb{Z}_2$ symmetry, while the untwisted sector is even. Among the scalar operators there are two with dimension $h_{1,7}= \frac{9}{7}$; one of them is odd and the other is even. Similarly, there are two operators with dimension $h_{2,7}= -\frac{4}{35}$. We identify the $\mathbb{Z}_2$ odd one, $\phi^{-}_{2, 7}$, with GL field $\phi$.  

In addition to the scalar primary operators, which are listed in Table \ref{tab:M514}, there are several primary operators with spin. 
There is a spin-$1$ operator $\phi_{2, 5} \bar\phi_{2, 9}\oplus \phi_{2, 9} \bar\phi_{2, 5}$ with dimensions $(-\frac{9}{35}, \frac{26}{35})$ and $(\frac{26}{35}, -\frac{9}{35})$;
a spin-$2$ operator $\phi_{2, 3} \bar\phi_{2, 11}\oplus \phi_{2, 11} \bar\phi_{2, 3}$ with dimensions $(\frac{11}{35}, \frac{81}{35})$ and $(\frac{81}{35}, \frac{11}{35})$;
a spin-$6$ operator $\phi_{1, 3} \bar\phi_{1, 11}\oplus \phi_{1, 11} \bar\phi_{1, 3}$ with dimensions $(-\frac{2}{7}, \frac{40}{7})$ and $(\frac{40}{7}, -\frac{2}{7})$;
a spin-$9$ operator $\phi_{1, 13} \oplus \bar\phi_{1, 13}$ with dimensions $(9, 0)$ and $(0, 9)$ (the spin-$9$ current is conserved in 2D). There are also two spin-$3$ operators:
$\phi_{1, 5} \bar\phi_{1, 9}\oplus \phi_{1, 9} \bar\phi_{1, 5}$ with dimensions $(\frac{1}{7}, \frac{22}{7})$ and $(\frac{22}{7}, \frac{1}{7})$, and
$\phi_{2, 1} \bar\phi_{2, 13}\oplus \phi_{2, 13} \bar\phi_{2, 1}$ with dimensions $(\frac{8}{5}, \frac{23}{5})$ and $(\frac{23}{5}, \frac{8}{5})$.

We propose the following identification for some of the primary operators of $M(5,14)$\footnote{We have checked some terms in the OPE, where in addition to the free field contractions we include the quintic interaction terms in the action (\ref{fiveaction}). However, our operator identifications should be regarded as tentative.} 
\begin{align}
& \phi_{1, 3}\sim \sigma \ , \quad \phi^{-}_{2, 7}\sim \phi, \quad
\phi_{1,5} \sim i\sigma^2+i\phi^2\ ,\quad \phi_{1, 7}^+\sim i\sigma\phi^4+i\sigma^3\phi^2+i\sigma^5\ , \\
& \phi_{2, 5} \bar\phi_{2, 9}\oplus \phi_{2, 9} \bar\phi_{2, 5} \sim \sigma \partial_\mu \phi- \phi \partial_\mu \sigma\ ,
\qquad \phi_{2, 3} \bar\phi_{2, 11}\oplus \phi_{2, 11} \bar\phi_{2, 3} \sim  \sigma \partial_\mu \partial_\nu \phi + \phi \partial_\mu \partial_\nu \sigma + \ldots \nonumber
\end{align}

\begin{table}[h!]
        \centering
          \begin{adjustbox}{max width=\textwidth}
          \renewcommand{\arraystretch}{1.2}
        \begin{tabular}{|c|c|c|c|c|c|c|c|c|c|c|c|c|c|}
            \hline
            $(5,14)$ & $\phi_{1,1}$ & $\phi_{1,3}$ & $\phi_{1,5}$ & $\phi_{1,7}^+$ & $\phi^{-}_{1,7}$ & $\phi_{1,9}$ & $\phi_{1,11}$ & $\phi_{1,13}$\\\hline
            $h$ & $0$ & $-\frac{2}{7}$ & $\frac{1}{7}$ & $\frac{9}{7}$ & $\frac{9}{7}$ & $\frac{22}{7}$ & $\frac{40}{7}$ & $9$ \\\hline
            $\mathbb{Z}_2$ & even & even & even & even & odd & even & even & even\\\hline
            $\mathcal{PT}$ & even & odd & odd & even &even & odd& odd & even \\\hline
            $(5,14)$ & $\phi_{2,1}$ & $\phi_{2,3}$ & $\phi_{2,5}$ & $\phi_{2,7}^+$ & $\phi^{-}_{2,7}$ & $\phi_{2,9}$& $\phi_{2,11}$ & $\phi_{2,13}$\\\hline
            $h$ & $\frac{8}{5}$ & $\frac{11}{35}$ & $-\frac{9}{35}$ & $-\frac{4}{35}$ & $-\frac{4}{35}$ & $\frac{26}{35}$ & $\frac{81}{35}$ & $\frac{23}{5}$ \\\hline
            $\mathbb{Z}_2$ & even & even & even & even & odd & even & even & even  \\\hline
            $\mathcal{PT}$ & even &odd & odd & even& even & odd & odd& even  \\\hline
        \end{tabular}
          \end{adjustbox}
        \caption{Spin-$0$ primary operators and their properties for the $D_8$ modular invariant of $M(5,14)$. We choose $\phi^{-}_{1,7}$ and $\phi^{-}_{2,7}$, which are from the twisted sector of the $\mathbb{Z}_2$ orbifold, to be $\mathcal{PT} $ even.
}
        \label{tab:M514}
    \end{table}
    \begin{table}[h!]
        \centering
          \begin{adjustbox}{max width=\textwidth}
          \renewcommand{\arraystretch}{1.2}
        \begin{tabular}{|c|c|c|c|c|c|c|c|c|c|c|c|c|c|}
            \hline
            $(5,16)$ & $\phi_{1,1}$ & $\phi_{1,3}$ & $\phi_{1,5}$ & $\phi_{1,7}$ & $\phi^-_{1,8}$ & $\phi_{1,9}$ & $\phi_{1,11}$ & $\phi_{1,13}$ & $\phi_{1,15}$ \\\hline
            $h$ & $0$ & $-\frac{3}{8}$ & $-\frac{1}{8}$ & $\frac{3}{4}$ & $\frac{91}{64}$ & $\frac{9}{4}$ & $\frac{35}{8}$ & $\frac{57}{8}$ & $\frac{21}{2}$ \\\hline
            $\mathbb{Z}_2$ & even & even & even & even & odd & even & even & even & even\\\hline
            $\mathcal{PT}$ & even & even & odd & even & even & even & odd & even & even\\\hline
            $(5,16)$ & $\phi_{2,1}$ & $\phi_{2,3}$ & $\phi_{2,5}$ & $\phi_{2,7}$ & $\phi^-_{2,8}$ & $\phi_{2,9}$& $\phi_{2,11}$ & $\phi_{2,13}$ & $\phi_{2,15}$\\\hline
            $h$ & $\frac{19}{10}$ & $\frac{21}{40}$ & $-\frac{9}{40}$ & $-\frac{7}{20}$ & $-\frac{57}{320}$ & $\frac{3}{20}$ & $\frac{51}{40}$ & $\frac{121}{40}$ & $\frac{27}{5}$ \\\hline
            $\mathbb{Z}_2$ & even & even & even & even & odd & even & even & even & even\\\hline
            $\mathcal{PT}$ & even & even & odd & even & even & even & odd & even & even \\\hline
        \end{tabular}
          \end{adjustbox}
        \caption{Spin-$0$ primary operators and their properties for the $D_9$ modular invariant of $M(5,16)$. We choose $\phi^-_{1,8}$ and $\phi^-_{2,8}$, 
which are from the twisted sector, to be $\mathcal{PT} $ even. }
        \label{tab:M516}
    \end{table}

The $D_9$ partition function of $M(5,16)$ is 
\begin{equation}
\begin{aligned}
    Z^{D_9}_{5,16}&=\sum_{i=1}^2(|\chi_{i,1}|^2+|\chi_{i,3}|^2+ |\chi_{i,5}|^2+|\chi_{i,7}|^2 +|\chi_{i,8}|^2+|\chi_{i,9}|^2 + |\chi_{i,11}|^2+|\chi_{i,13}|^2+\\
    &+|\chi_{i,15}|^2+\chi_{i,2}\bar\chi_{i,14}+\chi_{i,14}\bar\chi_{i,2}+\chi_{i,4}\bar\chi_{i,12}+\chi_{i,12}\bar\chi_{i,4}+\chi_{i,6}\bar\chi_{i,10}+\chi_{i,10}\bar\chi_{i,6})~,
\end{aligned}
\end{equation}
and the scalar primary operators are listed in Table \ref{tab:M516}. In addition, there are several primary operators with spin. 
There is a spin-$1$ operator $\phi_{2, 6} \bar\phi_{2, 10}\oplus \phi_{2, 10} \bar\phi_{2, 6}$ with dimensions $(-\frac{117}{320}, \frac{203}{320})$ and 
$(\frac{203}{320}, -\frac{117}{320})$; 
a spin-$2$ operator $\phi_{2, 4} \bar\phi_{2, 12}\oplus \phi_{2, 12} \bar\phi_{2, 4}$ with dimensions $(\frac{23}{320}, \frac{663}{320})$ and $(\frac{663}{320}, \frac{23}{320})$;
a spin-$6$ operator $\phi_{1, 4} \bar\phi_{1, 12}\oplus \phi_{1, 12} \bar\phi_{1, 4}$ with dimensions $(-\frac{21}{64}, \frac{363}{64})$ and $(\frac{363}{64}, -\frac{21}{64})$;
a spin-$9$ operator $\phi_{1,2} \bar \phi_{1, 14} \oplus \phi_{1, 14}\bar \phi_{1,2}$ with dimensions $(-\frac{17}{64}, \frac{559}{64})$ and $(\frac{559}{64}, -\frac{17}{64})$. 
There are also two spin-$3$ operators:
$\phi_{1, 6} \bar\phi_{1, 10}\oplus \phi_{1, 10} \bar\phi_{1, 6}$ with dimensions $(\frac{15}{64}, \frac{207}{64})$ and $(\frac{207}{64}, \frac{15}{64})$, and
$\phi_{2, 2} \bar\phi_{2, 14}\oplus \phi_{2, 14} \bar\phi_{2, 2}$ with dimensions $(\frac{363}{320}, \frac{1323}{320})$ and $(\frac{1323}{320}, \frac{363}{320})$.
We note that the spectrum of spins is the same as in $M(5,14)$, only the scaling dimensions are different.

Some explicit field identifications in $M(5,16)$ are 
\begin{align}
& \phi_{1, 3}\sim \sigma^2+ \phi^2 \ , \quad \phi^{-}_{2, 8}\sim \phi\ , \quad
\phi_{1,5} \sim \sigma\ ,\quad \phi_{1, 7}\sim i\sigma\phi^4+i\sigma^3\phi^2+i\sigma^5\ , \\
& \phi_{2, 6} \bar\phi_{2, 10}\oplus \phi_{2, 10} \bar\phi_{2, 6} \sim \sigma \partial_\mu \phi- \phi \partial_\mu \sigma\ , \qquad
 \phi_{2, 4} \bar\phi_{2, 12}\oplus \phi_{2, 12} \bar\phi_{2, 4} \sim  \sigma \partial_\mu \partial_\nu \phi + \phi \partial_\mu \partial_\nu \sigma + \ldots \nonumber
\end{align}

Let us look for the RG fixed points in $d=10/3-\epsilon$ that may correspond to $M(5,14)$ and $M(5,16)$.
The leading-order beta functions of the quintic action (\ref{fiveaction}) may be extracted from \cite{Gracey:2017okb}:
\small
        \begin{align}
        \beta_1&=-\frac{3}{2}g_1\epsilon\!-\!\frac{\Gamma(\frac{2}{3})^3}{80}(11889g_1^3+17280g_1^2g_2+13974g_1g_2^2+10800g_2^3-480g_1g_2g_3+1620g_2^2g_3-3g_1g_3^2),\nonumber\\
        \beta_2&=-\frac{3}{2}g_2\epsilon\!-\!\frac{\Gamma(\frac{2}{3})^3}{80}(2880g_1^3\!+\!6987g_1^2g_2\!+\!1620g_1g_2(10g_2\!+\!g_3)\!+\!17622g_2^3\!+\!120g_3(72g_2^2\!-\!g_1^2)\!+\!1251g_2g_3^2),\nonumber\\
        \beta_3&=-\frac{3}{2}g_3\epsilon\!-\!\frac{\Gamma(\frac{2}{3})^3}{16}(-240g_1^2g_2+1620 g_1g_2^2+5760g_2^3-3g_1^2g_3+2502g_2^2g_3+1377g_3^3).
        \end{align}
\normalsize
They may be obtained as gradients, 
\begin{equation}
\beta_1 = \frac{\partial C}{\partial g_1}\ , \qquad \beta_2 = \frac{1}{2} \frac{\partial C}{\partial g_2}\ , \qquad \beta_3 = 5 \frac{\partial C}{\partial g_3}\ ,
\end{equation}
and the $C$-function is given by
\small
\begin{align}
& C(g_1, g_2, g_3)=  -\epsilon \left ( \frac{3}{4} g_1^2 +  \frac{3}{2} g_2^2 + \frac{3}{20} g_3^2 \right )
 \!-\!\frac{\Gamma(\frac{2}{3})^3}{80}
\bigg ( \frac{11889}{4} g_1^4+5760 g_1^3 g_2+ 6987 g_1^2g_2^2+ 10800 g_1 g_2^3 \nonumber \\
&  -240 g_1^2 g_2g_3+1620 g_1 g_2^2g_3-\frac{3}{2} g_1^2 g_3^2 + 8811 g_2^4 
+5760 g_2^3 g_3 + 1251 g_2^2 g_3^2 + \frac{1377}{4} g_3^4
\bigg )\ .
\end{align}    
\normalsize
The beta functions have four purely imaginary fixed points. The fixed point with $g_1^*=g_2^*=0$ describes the $M(2, 7)$ model together with a free field. The fixed point with $g_1^*=g_2^*=g_3^*$ corresponds to two decoupled copies of $M(2, 7)$, because the interaction is proportional to  the sum of  $(\sigma\pm\phi)^5$ at this fixed point. 
There are two more nontrivial fixed points:
\begin{align}
& (g^*_1,g^*_2,g^*_3)_{\rm I} = i\frac{\sqrt{\epsilon}}{3\sqrt{163}}  \Gamma\left(\frac{2}{3}\right)^{-\frac{3}{2}}\left(3, 1, -5\right)
\approx i \sqrt{\epsilon}\, \Gamma\left(\frac{2}{3}\right)^{-\frac{3}{2}}\left(0.0783, 0.0261, -0.1305\right)
~, \nonumber \\
& (g^*_1,g^*_2,g^*_3)_{\rm II} \approx i \sqrt{\epsilon}\, \Gamma\left(\frac{2}{3}\right)^{-\frac{3}{2}}\left(0.0397, 0.0484, 0.0591\right)~.
\end{align}
The stability matrix $\partial_{g_i} \beta_j $ has eigenvalues $(3.008\epsilon, 3\epsilon, -1.107\epsilon)$ and  $(3\epsilon, 0.0065\epsilon, -1.120\epsilon)$ at these two fixed points, respectively. We propose that fixed point I, where $C\approx 0.001647 \epsilon^2$, corresponds to the $M(5, 14)$ model, while fixed point II, where 
$C\approx 0.001053 \epsilon^2$, corresponds to the $M(5,16)$ model. 
The flow from the latter to the former would then correspond to (\ref{threeflow}) with $q=5$. We note that $C_{\rm IR} > C_{\rm UV}$, but this is typical for RG flows connecting non-unitary fixed points where the coupling constants are imaginary.\footnote{Let us also note that the two fixed points have the same number of relevant and irrelevant operators, while typically one of the relevant operators in the UV turns into an irrelevant one in the IR. However, one of the irrelevant operators at the UV fixed point is very close to being marginal. So, it seems possible that it turns into a relevant one during the RG flow, since the IR and UV fixed points are not close to each other. We leave a better understanding of this issue for the future.}

Our proposed GL description of $M(7, 20)$ and $M(7,22)$ is in terms of the field theory with interactions of seventh order:
    \begin{align}
        S_{7}=\int d^dx\left(\frac{1}{2}(\partial_\mu\phi)^2+\frac{1}{2}(\partial_\mu\sigma)^2+\frac{g_1}{6!}\sigma\phi^6+\frac{g_2}{4!\cdot3!}\sigma^3\phi^4+\frac{g_3}{2!\cdot 5!}\sigma^5\phi^2+\frac{g_4}{7!}\sigma^7\right)~.
    \end{align}
Using the one-loop beta functions in $14/5-\epsilon$ dimensions \cite{Gracey:2017okb}, we find three nontrivial purely imaginary fixed points  in addition to two more that correspond to decoupled theories: 
\begin{align}
    &(g_1^*, g_2^*, g_3^*, g_4^*) = \frac{3 i\sqrt{\epsilon}(5, -1, -3, 7)}{\Gamma \left(\frac{4}{5}\right)^2 \sqrt{14400 \Gamma \left(\frac{3}{5}\right)^2+23410 \Gamma \left(\frac{4}{5}\right)}},\nonumber\\
    &(g_1^*, g_2^*, g_3^*, g_4^*) = \frac{3 i \sqrt{\epsilon }(9, 3, 1, -21)}{\Gamma \left(\frac{4}{5}\right)^2 \sqrt{172800 \Gamma \left(\frac{3}{5}\right)^2+233398 \Gamma \left(\frac{4}{5}\right)}},\\
    &(g_1^*, g_2^*, g_3^*, g_4^*) \approx i\sqrt{\epsilon}(0.006656, 0.008995, 0.012156 ,0.016431)~.\nonumber
\end{align}
The first fixed point has one relevant direction while the other two have two relevant directions.  We expect the first fixed point to describe the $M(7,20)$ minimal model because it is the most stable. We plan to consider this theory in more detail in the future.

\section*{Acknowledgments}

We are grateful to S. Giombi, J. Maldacena, A. Miscioscia, Y. Nakayama, V. Narovlansky, A.M. Polyakov, N. Seiberg and G. Tarnopolsky for very useful discussions. 
We also thank the referee for useful questions and comments.
This work was supported in part by the US National Science Foundation Grant No.~PHY-2209997 and by the Simons Foundation Grant No.~917464.


\bibliographystyle{ssg}
\bibliography{GLsome}

\end{document}